\documentclass{article}
\usepackage{authblk}

\usepackage{etex}

\usepackage{blindtext}
\usepackage{array}

\usepackage{pdfpages}
\usepackage[noadjust]{cite}

\usepackage{url, graphicx, color, varioref, amsfonts, longtable, float,setspace,amsthm}
\usepackage[compact]{titlesec}
\usepackage{amsmath}
\usepackage{wrapfig}
\usepackage{pgfplots}
\usepackage{enumerate}
\usepackage{amssymb}
\usepackage{algorithm}
\usepackage{tikz}
\usepackage{mathtools}
\usepackage{epsfig}
\input{Qcircuit}
\usepackage[noend]{algpseudocode}

\usepackage[justification=centering]{caption}
\numberwithin{equation}{section}
\usepackage{hyperref}
\theoremstyle{definition}
\newtheorem{definition}{Definition}[section]

\title{Measuring Hamming Distance between Boolean Functions via Entanglement Measure}
\author {Khaled El-Wazan \thanks{khaled\_ elwazan@alex-sci.edu.eg} }

\affil{Department of Mathematics and Computer Science, Faculty of Science, Alexandria University, Egypt}

\date{}

\begin{document}
\maketitle

\begin{abstract}
In this paper, we present a fast quantum algorithm to measure the Hamming distance between two or more Boolean functions provided as black-boxes. The proposed algorithm constructs a new black-box with a certain property which is utilized to solve this problem. The introduced algorithm converts measuring the Hamming distance between Boolean functions to measuring entanglement between qubits, using concurrence entanglement measure.
\end{abstract}

\section{Introduction}
Given two Boolean functions $f$ and $g$ both with $n$ inputs, the Hamming distance between $f$ and $g$ is defined as the number of input vectors that outputs different results for both $f$ and $g$ \cite{Galatenko2011,wu2016boolean}.
%, \textit{i.e.} $H(f,g)=\vert\{ x\in\{0,1\}^n:f(x)\neq g(x)\}\vert$. 
To solve this problem classically with exact solution, one will iterate over all possible $N=2^n$ inputs for both Boolean functions and count when the output of $f$ and $g$ differs when the input is the same. This procedure will require $\mathcal{O}(N)$. When generalizing the problem of measuring the Hamming distance of $\kappa$ Boolean functions, the problem will require $\mathcal{O}(\kappa N)$.

Quantum computers \cite{Feynman1986,loyed} are promising probabilistic devices that guarantee to solve some problems faster than classical computers. %, by the utilization of quantum phenomena like entanglement and superposition. 
For example, Deutsch and Jozsa introduced a quantum algorithm \cite{deutsch-jozsa} to decide whether a given black-box with $n$ inputs represents a constant or balanced Boolean function. L. Grover provided an optimal quantum algorithm \cite{grover,grover-optimal} to find a single item in an unstructured list of $N$ items, only using $\mathcal{O}(\sqrt{N})$ oracle calls, which was later generalized by Boyer \textit{et al.} \cite{Boyer1996} to search for $M$ items using $\mathcal{O}(\sqrt{N/M})$ oracle calls.

Having a fast quantum subroutine to decide the Hamming distance between Boolean functions is useful. For example, it can be used as a preliminary test before attempting to solve a system of binary multivariate equations via Grover algorithm   \cite{schwabe2016solving}, or to find common matches between databases using quantum search algorithm with partial diffusion \cite{El-Wazan2017-CDBE,younes-miller}.

%Computation learning theory is the study of reconstruction of a Boolean function $f:\{0,1\}^n\rightarrow\{0,1\}$ with black-box access, with as minimum black-box queries as possible. It is useful to have a subroutine that can measure the similarities between the reconstructed black-box and the given black-box. 
%It is necessary useful as well to measure the Hamming distance between 

In 2018, Xie \textit{et al.} proposed a quantum algorithm \cite{Xie2018} based on Bernstein-Vazirani algorithm \cite{Bernstein-Vazirani} to measure the Hamming distance between two Boolean functions with $n$ inputs that requires $\mathcal{O}(1)$ in some cases with success probability at least $8/\pi^2$. As well, Xie \textit{et al. } proposed a quantum algorithm \cite{Xie2018}, based on quantum amplitude amplification and estimation algorithm \cite{Grover1997,Boyer1996}, that measures the Hamming distance between two Boolean functions given that the Hamming distance is $t$ $(t\neq 0)$, and requires $\mathcal{\theta}\big(\sqrt{\frac{N}{\lfloor \epsilon t\rfloor +1}}+\sqrt{\frac{t(N-1)}{\lfloor \epsilon t\rfloor +1}}\big)$ queries with accuracy $\epsilon$.

Quantum entanglement \cite{EPR,mintert2004concurrence,chen2005concurrence,zhang2007optimal} is one of the quantum phenomena that established itself as a crucial and useful resource for processing quantum information and quantum communication \cite{neilson-chuang,kaye2007introduction}. For example, it is utilized in quantum search algorithm with reliable behavior \cite{younes-miller}, quantum junta testing and learning of Boolean functions \cite{El-Wazan2017} and quantum key distribution \cite{liao2017satellite}. Many entanglement-based applications \cite{horodecki2009quantum} require detection of such phenomenon and quantifying it. Several methods of entanglement detection have been proposed \cite{e-detection} such as entanglement witness operator \cite{arrazola2012accessible,jungnitsch2011taming}, quantum state tomography \cite{gross2010quantum,cramer2010efficient} and concurrence entanglement measure \cite{walborn,singh2018experimentally}.

In this paper, we propose a fast quantum algorithm to measure the Hamming distance between two Boolean functions provided as black-boxes. The proposed algorithm utilizes quantum superposition to mark the common inputs that satisfy both the black-boxes with entanglement. The suggested algorithm converts the problem of measuring the Hamming distance between Boolean functions to measuring entanglement between qubits. The proposed algorithm is later generalized to measure the Hamming distance of $\kappa$ Boolean functions provided as black-boxes. The introduced algorithm works even if the Hamming distance is equal to zero.

This paper is organized as follows: 
Section ~\ref{preliminaries} introduces the basics and concepts of quantum entanglement and concurrence entanglement measure. 
Section ~\ref{construction} depicts the construction of the new black-box.
Section ~\ref{the-proposed-algorithm} introduces the proposed quantum algorithm.
Section ~\ref{analysis} provides analysis of the proposed algorithm, followed by a conclusion in Section ~\ref{conclusion}.

\section{Preliminaries}
\label{preliminaries}
\subsection{Notations and Definitions}

%\begin{definition} Given two Boolean functions $f$ and $g$ both with $n$ inputs, the Hamming distance between $f$ and $g$ is defined as the number of vectors of size $n$ when the output for both $f$ and $g$ differs \cite{Xie2018,Galatenko2011,Wu}, \textit{i.e.}
%
%\begin{equation}
%H(f,g)=\vert\{ x\in\{0,1\}^n:f(x)\neq g(x)\}\vert.
%\end{equation}
%\end{definition}

\begin{definition}
Given a set of $\kappa\geq 2$ Boolean functions each with $n$ inputs, we say that the Hamming distance between the given Boolean functions is defined as the number of entries that maps those Boolean functions to different outputs, \textit{i.e.}
\begin{equation}
H(f_0,f_1,\cdots,f_{\kappa-1})=\vert ~ x\in\{0,1\}^n: f_0(x)\neq f_1(x)\neq\cdots \neq f_{\kappa-1}(x)\vert.
\end{equation}

\end{definition}

\begin{definition}
We say that the pure state $\ket{\psi}$ in Hilbert space $\mathcal{H}$ is separable, if we can decompose the state $\ket{\psi}$ to the states $\ket{\phi_A}$ and $\ket{\phi_B}$ such that:
\begin{equation}
\ket{\psi}=\ket{\phi_A}\otimes \ket{\phi_B},
\end{equation}
otherwise, the state $\ket{\psi}$ is called entangled \cite{e-detection}.

\end{definition}

\begin{definition}
Toffoli gate \cite{toffoli} is a gate with three qubits, taking three qubits as input and producing the result by flipping the third qubit, which is considered the target qubit, if and only if the first two qubits are equal to $\ket{1}$. Figure ~\ref{fig:toffoli} illustrates Toffoli gate.

\begin{figure}[H]
\begin{align*}
 \Qcircuit @C=1em @R=.7em {
  \lstick{\ket{a}} &\qw & \ctrl{2}& \qw &\rstick{\ket{a}}\\
   \lstick{\ket{b}} &\qw & \ctrl{1}& \qw &\rstick{\ket{b}}\\
    \lstick{\ket{c}} &\qw & \targ & \qw &\rstick{\ket{c^\prime}}\\  
  }
\end{align*}
\caption{Toffoli gate where $\bullet$ represents the control qubit and  $\oplus$ marks the  target qubit, and $c^\prime=c\oplus a\cdot b$ where $\oplus$ is the classical \textit{XOR} operation. \label{fig:toffoli}}
\end{figure}
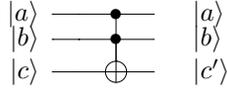

\end{definition}

\begin{definition}
For a given Boolean function $z:\{0,1\}^n\rightarrow\{0,1\}$, we define unitary gate $U_z$ such that:
\begin{equation}
U_z\ket{x}\otimes \ket{0}=\ket{x}\otimes \ket{z(x)},
\end{equation}
where $x\in\{0,1\}^n$.
 
\end{definition}

\begin{definition}
Given a black-box  $U_z$ representing a Boolean function $z$, we state that the black-box  $U_z$ is a unitary operator working on $n+t+q+1$ qubits, taking the control from the first $n$ qubits, $0\rightarrow n-1$, and its target is the qubit indexed $n+t$ \cite{El-Wazan2017-CDBE}. The stated configuration is denoted as $^{0\rightarrow n-1}_{n+t}U_z$. Figure ~\ref{fig:operator+target} illustrates the defined black-box.

\begin{figure}[H]
\begin{align*}
\Qcircuit @C=1em @R=1em { 
 \lstick{\ket{x_0}} & \qw  & \qw  & \multigate{3}{U_z}& \qw  & \qw  & \qw \\
 \lstick{\ket{x_1}} & \qw  & \qw  & \ghost{U_z}& \qw  & \qw   & \qw\\ 
  &  \vdots &  &&& \vdots\\ 
 \lstick{\ket{x_{n-1}}} & \qw  & \qw  & \ghost{U_z}& \qw  & \qw  & \qw \\ 
 \lstick{\ket{x_n}} & \qw  & \qw  & \qwx \qw  & \qw  & \qw   & \qw\\ 
 &  \vdots & &\qwx && \vdots\\ 
 \lstick{\ket{x_{n+t-1}}} & \qw  & \qw  & \qwx  \qw   & \qw  & \qw  & \qw \\ 
 \lstick{\ket{x_{n+t}}}& \qw  & \qw  & \targ \qwx &\qw  & \qw  & \qw  \\ 
 \lstick{\ket{x_{n+t+1}}}& \qw  & \qw  & \qw  & \qw  & \qw   & \qw \\ 
 &  \vdots & &&& \vdots\\ 
 \lstick{\ket{x_{n+t+q}}}& \qw  & \qw  & \qw  & \qw  & \qw  & \qw \\ 
}
\end{align*}
\caption{ A quantum circuit representing the black-box $\strut^{0\rightarrow n-1}_{n+t}U_z$ \cite{El-Wazan2017-CDBE}.\label{fig:operator+target}}
\end{figure}
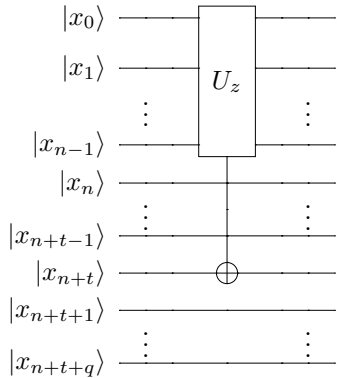

\end{definition}

\subsection{Black-Box Categorization Algorithm}
\begin{figure}[H]
\begin{align*}
\Qcircuit @C=1em @R=1em { 
 \lstick{\ket{x_0}} & \qw  & \qw  & \qw& \qw  & \qw  & \qw & \qw  & \qw\\
 \lstick{\ket{x_1}} & \qw  & \qw  & \qw& \qw  & \qw   & \qw & \qw& \qw\\ 
  &  \vdots &  &&& \vdots &  \\ 
\lstick{\ket{x_i}} & \qw & \qw & \ctrl{1}& \qw  & \qw  & \qw & \gate{D}\cwx[2]& \qw\\
   & \vdots &  &&& \vdots & \\ 
 \lstick{\ket{x_{n-1}}} & \qw  & \qw& \qw\qwx & \qw &\qw& \qw  & \qw& \qw \\ 
 \lstick{\ket{x_n}} &\qw & \qw & \targ\qwx & \qw &\qw & \qw & \gate{D}\cwx[-1]& \qw
}
\end{align*}
\caption{A quantum circuit representing the proposed operator $U_\lambda$ \cite{El-Wazan2017}.\label{Ul}}
\end{figure}
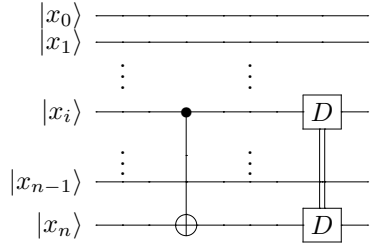

In \cite{El-Wazan2017}, we introduced a quantum algorithm that uses concurrence entanglement measurement operator to categorize any given Black-box $U_f$ representing a Boolean function $f$ to either a constant, a balanced or a Boolean function of other form. The proposed quantum entanglement measurement operator $U^i_\lambda$ \cite{El-Wazan2017} acts on a given qubit with index $i$ and an extra qubit initialized with the state $\ket{1}$, and creates a measurable entanglement between the qubit indexed $i$ and the extra qubit using \textit{CNOT} gate. Figure ~\ref{Ul} represents the proposed operator,

where $D$ is an entanglement measure device that measures the entanglement between the designated qubits $\ket{x_i}$ and the extra qubit $\ket{x_n}$.

The entanglement will happen if and only if the qubit indexed $i$ is in superposition, and for a quantum system that exhibits entanglement,

\begin{equation}
\ket{\psi}=\alpha\ket{01}+\beta\ket{01},
\end{equation}
the concurrence is defined as follows \cite{walborn}:
\begin{equation}
C(\ket{\psi})=\vert 2\alpha\beta\vert.
\end{equation}

The algorithm starts with the initial state $\ket{0}^{n+1}\otimes \ket{1}$ and proceeds as follows:

\begin{align}
&\ket{0}^{n+1}\otimes \ket{1}  \\
&\underrightarrow{H^{\otimes n}\otimes I^{\otimes 2}} \frac{1}{\sqrt{N}} \sum_{l=0}^{N-1}\ket{l}\otimes \ket{0}\otimes \ket{1} \\
&\underrightarrow{\strut^{0\rightarrow n-1}_{n}U_f} \Big(\frac{1}{\sqrt{N}} \sum_{l=0}^{N-1}\strut^{\prime\prime} \ket{l}\otimes \ket{0}+\frac{1}{\sqrt{N}}\sum_{l=0}^{N-1}\strut^{\prime} \ket{l}\otimes \ket{1}\Big)\otimes \ket{1} \\
&\underrightarrow{H^{\otimes n}\otimes I^{\otimes 2}} \Big(\sum_{l=0}^{N-1}\tilde{f}_0(l)\ket{l}\otimes \ket{0} + \sum_{l=0}^{N-1}\tilde{f}_1(l)\ket{l}\otimes \ket{1}\Big)\otimes \ket{1}, 
%&\underrightarrow{U^n_\lambda}\alpha_n\ket{01}+\beta_n\ket{10},
\end{align}
where $\tilde{f}_0$ and $\tilde{f}_1$ are 

\begin{align}
\tilde{f}_0(l)&=\frac{1}{N}\sum_{s=0}^{N-1}\strut^{\prime\prime}(-1)^{l\cdot s}, \\
\tilde{f}_1(l)&=\frac{1}{N}\sum_{s=0}^{N-1}\strut^{\prime}(-1)^{l\cdot s}.
\end{align}

By applying the concurrence measurement operator $U^n_\lambda$ on the last two qubits, the measured concurrence can be expressed as follows \cite{El-Wazan2017,zidan2018novel}:
\begin{equation}
C=2\times \frac{\sqrt{M(N-M)}}{N},
\label{eq:Conc}
\end{equation}
where $M$ is the number of solutions that satisfies the black-box $U_f$ such that  $0\leq M\leq N$. Depending on the measured concurrence, the Boolean function $f$ is categorized as follows:
\begin{equation}
C=\begin{cases}
0, \textit{constant Boolean function}\\
1/2, \textit{balanced Boolean function}\\
\textit{Otherwise, Boolean function of other form}
\end{cases}.
\end{equation}

\section{Constructing the Black-box $U_\kappa$}
\label{construction}
In this section, given $\kappa\geq 2$ black-boxes representing Boolean functions with $n$ inputs,  we will construct a new black-box $U_\kappa$ which will be utilized to measure the Hamming distance between those Boolean functions. For illustrative purposes, we will construct $U_\kappa$ for $\kappa=2$ black-boxes, and after this, we will broaden the constructed  black-box $U_\kappa$ for $\kappa\geq 2$.

\subsection{Constructing $U_\kappa$ for two Boolean Functions}
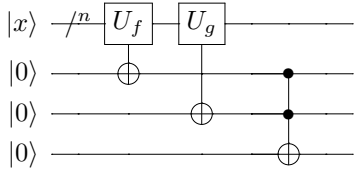
\begin{figure}[H]
\begin{align*}
\Qcircuit @C=1em @R=.7em {
  \lstick{\ket{x}} &  /^n \qw  & \gate{ U_f}& \gate{U_g} & \qw	&\qw  & \qw & \qw\\
  \lstick{\ket{0}} & \qw      & \targ \qwx &\qw\qwx& \qw  & \ctrl{1}\qw  & \qw& \qw \\  
  \lstick{\ket{0}} & \qw       & \qw &\targ\qwx& \qw &\ctrl{1}\qw    & \qw & \qw  \\    
 \lstick{\ket{0}} &\qw& \qw & \qw & \qw  &\targ \qw & \qw & \qw \\
 %\lstick{\ket{1}} & \qw & \qw & \qw & \qw  & \qw & \qw & \ghost{U_\lambda}\\
 }
\end{align*}
\caption{The constructed black-box $U_\kappa$ for $\kappa=2$ black-boxes.\label{fig:proposed-alg-k-2}}
\end{figure}

In this section, we construct a new black-box from two black-boxes $U_f$ and $U_g$ representing Boolean functions $f$ and $g$, respectively. Figure ~\ref{fig:proposed-alg-k-2} illustrates the proposed black-box.

We can represent the constructed black-box $U_\kappa$ for $\kappa=2$ as a quantum circuit defined as follows:
\begin{equation}
U_\kappa=  \strut^{n\rightarrow n+1} _{n+2}T\times \strut^{0\rightarrow n-1}_{n+1}U_g \times \strut^{0\rightarrow n-1}_{n}U_f,
\end{equation}
where $\strut^{n\rightarrow n+1} _{n+2}T$ is the toffoli gate with two control qubits indexed $n$ and $n+1$, and target qubit indexed $n+2$.

Let's study the constructed black-box in the presence of uniform superposition.

\begin{enumerate}

\item \textit{Register Preparation.} Prepare a quantum register of $n+3$ qubits in the state $\ket{0}$:\label{fstep}

\begin{equation}
\vert\varphi_0\rangle=\vert 0\rangle^{\otimes n}\otimes \vert 0\rangle^{\otimes 3}.
\end{equation}

\item \textit{Register Initialization.} Apply Hadamard gate on the first $n$ qubits to get a uniform superposition of all the possible $N=2^n$ states:

\begin{align}
\vert\varphi_1\rangle&=H^{\otimes n}\vert\varphi_0\rangle \nonumber \\
&=H^{\otimes n}\vert 0\rangle^{\otimes n} \otimes \vert 0\rangle^{\otimes 3}\nonumber \\
&=\frac{1}{\sqrt{N}}\sum_{l=0}^{N-1}\vert l\rangle \otimes \vert 0\rangle^{\otimes 3}.
\end{align}

\item \textit{Applying the Oracle $U_f$.} Apply the oracle $U_f$ on the register to mark all possible solutions of the function $f$ using the qubit indexed $n+1$:
%, where non-solutions will be marked with $\vert 0\rangle$ and the solutions will be marked with $\vert 1\rangle$: 

\begin{align}
\vert \varphi_2\rangle&=\strut^{0\rightarrow n-1} _{n}U_f \vert\varphi_1\rangle \nonumber \\
&=\frac{1}{\sqrt{N}}\sum_{l=0}^{N-1}\vert l\rangle \otimes \vert f(l)\rangle \otimes \vert 0\rangle^{\otimes 2}.
\end{align}

\item \textit{Applying the Oracle $U_g$.} Apply the oracle $U_g$ on the register to mark all possible solutions of the function $g$ using the qubit indexed $n+2$:
%, where the non-solutions will be marked with $\vert 0\rangle$ and the solutions will be marked with $\vert 1\rangle$: 

\begin{align}
\vert \varphi_3\rangle&=\strut^{0\rightarrow n-1} _{n+1}U_g \vert\varphi_2\rangle \nonumber \\
&=\frac{1}{\sqrt{N}}\sum_{l=0}^{N-1}\vert l\rangle \otimes \vert f(l)\rangle \otimes \vert g(l)\rangle \otimes \vert 0\rangle.
\end{align}

\item \textit{Applying the Toffloi gate.} Apply the toffloi gate on the qubits indexed $n+1$ and $n+2$ to mark all possible common solutions between the functions $f$ and $g$ using the qubit indexed $n+2$ as the target qubit, where non-common solutions will be marked with $\vert 0\rangle$ and the common solutions will be marked with $\vert 1\rangle$: 

\begin{align}
\vert \varphi_4\rangle&=\strut^{n\rightarrow n+1} _{n+2}T \vert\varphi_3\rangle \nonumber \\
&=\frac{1}{\sqrt{N}}\sum_{l=0}^{N-1}\vert l\rangle \otimes \vert f(l)\rangle \otimes \vert g(l)\rangle \otimes \vert f(l)\cdot g(l)\rangle ,
\end{align}
such that $\cdot$ is the AND logic operation.

\end{enumerate}

It is clear that after applying the black-box $U_\kappa$, we will have all the joint states that satisfy both the Boolean function $f$ and $g$ marked $\ket{1}$ using the qubit indexed $n+2$.

\subsection{Constructing $U_\kappa$ for $\kappa$ Boolean Functions}
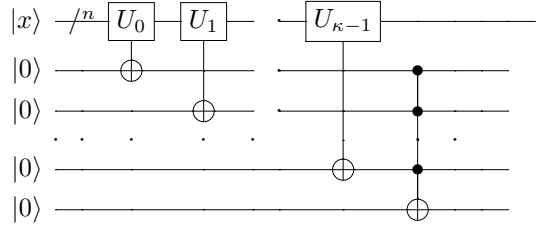
\begin{figure}[H]
\begin{align*}
\Qcircuit @C=1em @R=.7em {
  \lstick{\ket{x}} &  /^n \qw & \gate{ U_0}& \gate{U_1}&\qw & \cdot  & \gate{U_{\kappa-1}} &\qw& \qw& \qw&  \qw&  \qw	\\
  \lstick{\ket{0}}      & \qw  & \targ \qwx &\qw\qwx&\qw & \cdot & \qw\qwx  & \ctrl{1} & \qw&  \qw&  \qw\\  
  \lstick{\ket{0}}     & \qw  & \qw &\targ\qwx&\qw & \cdot     & \qw\qwx & \ctrl{1} \qwx& \qw &  \qw&  \qw \\
  \cdot & \cdot  & \cdot & \cdot & \cdot & &\qwx &  \cdot & \cdot & \\    
 \lstick{\ket{0}}  & \qw & \qw & \qw & \qw  &\qw& \targ\qwx  &\ctrl{1}\qwx& \qw&  \qw &  \qw\\
 \lstick{\ket{0}} & \qw & \qw & \qw & \qw  & \qw &\qw&  \targ\qwx & \qw & \qw&  \qw\\
 %\lstick{\ket{1}}& \qw & \qw & \qw & \qw  & \qw &\qw&  \qw & \qw&\ghost{U_\lambda}&  \qw\\
 }
\end{align*}
\caption{The constructed black-box $U_\kappa$ for $\kappa\geq 2$ black-boxes.\label{fig:general-alg-k-inf}}
\end{figure}

Given $\kappa\geq 2$ black-boxes all of $n$ inputs and $\kappa+1$ axillary qubits, we generalize the constructed black-box to measure the Hamming distance between the given $\kappa$ black-boxes. Figure ~\ref{fig:general-alg-k-inf} illustrates the proposed black-box.

We can represent the constructed black-box $U_\kappa$ for $\kappa\geq 2$ as a quantum circuit defined as follows:
\begin{equation}
U_\hbar= \strut^{n\rightarrow n+\kappa-1} _{n+\kappa}T \times \prod_{j=0}^{\kappa-1}\strut^{0\rightarrow n-1} _{n+j} U_j,
\end{equation}
where $\strut^{n\rightarrow n+\kappa-1} _{n+\kappa}T$ is the toffoli gate with $\kappa$ controls.

The general system in a uniform superposition for $\kappa \geq 2$ can be generally described as follows:

\begin{equation}
\ket{\psi}=\frac{1}{\sqrt{N}}\sum_{l=0}^{N-1}\ket{l}\bigotimes_{j=0}^{\kappa-1}f_j(l)\otimes  \ket{ \chi },
\end{equation}
where $\chi=\bigwedge_{j=0}^{\kappa-1}f_j(l)$ and $\wedge$ is the \textit{AND} logic operation.

\section{The Proposed Algorithm}
\label{the-proposed-algorithm}
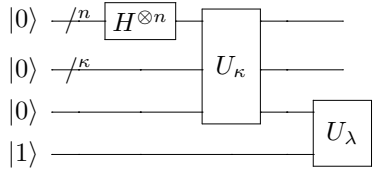
\begin{figure}[H]
\begin{align*}
\Qcircuit @C=1em @R=.7em {
  \lstick{\ket{0}} & /^n \qw & \gate{H^{\otimes n}} & \multigate{2}{U_{\kappa}} & \qw & \qw & \\
  \lstick{\ket{0}}   & /^\kappa  \qw  & \qw &  \ghost{U_\kappa} & \qw & \qw\\
   \lstick{\ket{0}}   & \qw  & \qw   &  \ghost{U_\kappa}& \qw& \multigate{1}{U_\lambda}   & \\
   \lstick{\ket{1}} & \qw & \qw & \qw & \qw & \ghost{U_\lambda}  & \\
 }
\end{align*}
\caption{Quantum circuit for the proposed algorithm.
\label{fig:the-proposed-alg}}
\end{figure}

In this section, we propose a quantum algorithm to measure the Hamming distance of $\kappa$ Boolean functions provided as black-boxes. The proposed algorithm utilizes the new constructed  black-box in Section ~\ref{construction} to measure the Hamming distance using concurrence entanglement measurement operator. Figure ~\ref{fig:the-proposed-alg} illustrates the proposed algorithm.

The algorithm is carried quantum mechanically as follows:

\begin{algorithm}[H]
\label{1st-algo}
\begin{algorithmic}[1]
\State Construct the oracle $U_\kappa$.
\State  Set the quantum register to $\ket{0}^{\otimes n}$ and the extra $\kappa+2$ qubits to $\ket{0}^{\otimes{\kappa+1}}\otimes \ket{1}$.
\State Apply the Hadamard gates to the first $n$ qubits to create the uniform superposition:
\begin{equation*}
\ket{\psi}=\frac{1}{\sqrt{N}}\sum_{l=0}^{N-1}{\ket{l}}\otimes \ket{0}^{\otimes \kappa+1}\otimes \ket{1}.
\end{equation*}

\State Apply the constructed black-box $U_\kappa$.
\State Apply the concurrence entanglement measurement operator $U_\lambda$, assuming the measured concurrence is $C$.

\If{$C\neq 0$} 
	\State \textbf{exit}.
\EndIf
\State Repeat the algorithm without applying the operator $U_\lambda$.
\State Measure the qubit indexed $n+\kappa+1$. \label{extra-step}

\end{algorithmic}
\caption*{The Proposed Algorithm.}
\end{algorithm}

\section{Analysis of the Proposed Algorithm}
\label{analysis}
In this section, we will elaborate the behavior of the proposed algorithm with the proposed concurrence entanglement measurement operator, with respect to all possible scenarios, for any given $\kappa$ black-boxes.

Let's assume the state of the quantum system 
%before applying the entanglement measurement operator $D$ 
in the proposed quantum algorithm before applying $U_\lambda$ and focusing on the last two qubits, indexed $n+\kappa+1$ and $n+\kappa+2$, is as follows:
\begin{equation}
\ket{\eta_0}=\big(\alpha\ket{0}+\beta\ket{1}\big)\otimes \ket{1}.
\end{equation}

Applying the \textit{CNOT} gate on $\ket{\eta_0}$:
\begin{align}
\ket{\eta_1}&=CNOT\ket{\eta_0} \nonumber \\
&=\alpha\ket{0, 1 \oplus CNOT(0) }+\beta\ket{1, 1\oplus CNOT(1)},
\end{align}
and after applying the quantum entanglement measurement operator $D$ on the quantum system $\ket{\eta_1}$, Equation ~\ref{eq:Conc} can be reformulated as follows:
\begin{equation}
C(\ket{\eta_1})=2\times \frac{\sqrt{M_c(N-M_c)}}{N},
\label{eq:the-mc}
\end{equation}
where $M_c$ is the number of common solutions between the given Boolean functions.

\subsection{In Case Concurrence is Detected  $(C\neq 0)$} In such case, the quantum system $\ket{\eta_1}$ can be described as follows:
\begin{equation}
\ket{\eta_1}=\alpha\ket{10}+\beta\ket{01},
\end{equation}
then after applying the operator $U_\lambda$, there will be a measurable entanglement. Solving Equation ~\ref{eq:the-mc} for unknown $M_c$, the Hamming distance between the Boolean functions can be defined as follows:
\begin{equation}
H(f_0,f_1,\cdots,f_{\kappa-1})=N-M_c.
\end{equation}

\subsection{In Case No Concurrence is Detected  $(C=0)$}  In this case, there are two reasons that will produce no entanglement:
 
\begin{enumerate}
 
\item When $M_c=0$, which means that all given Boolean functions produce different output when they are given the same input. In such case, the quantum system $\ket{\eta_1}$ can be described as follows:
\begin{equation}
\ket{\eta_1}=\ket{01}.
\end{equation}

\item When $M_c=N$, which means that all given Boolean functions agree on the output when the input is the same. In such case, the quantum system $\ket{\eta_1}$ can be described as follows:
\begin{equation}
\ket{\eta_1}=\ket{10}.
\end{equation}

\end{enumerate}

To be able to distinguish between those cases, we measure the qubit indexed $n+\kappa+1$ as in Step ~\ref{extra-step} of the proposed algorithm, assuming the output is $\delta$. If $\delta=\ket{0}$ then,
\begin{equation}
H(f_0,f_1,\cdots,f_{\kappa-1})=N,
\end{equation}
but if $\delta=\ket{1}$ then,
\begin{equation}
H(f_0,f_1,\cdots,f_{\kappa-1})=0.
\end{equation}

\section{Conclusion}
\label{conclusion}
In this paper, we introduced a fast quantum algorithm to measure the Hamming distance between Boolean functions provided as black-boxes.
Using the provides black-boxes, we constructed a new black-box that exhibits the behavior of finding the common solutions of the provided black-boxes.  
We transformed the problem of determining the Hamming distance between Boolean functions to measuring entanglement between qubits. 
The proposed quantum algorithm requires a single oracle call to all Boolean functions to determine the Hamming distance, and works even when the Hamming distance is equal to zero, opposite to relevant work.

\bibliographystyle{IEEEtran}
\bibliography{references}

\end{document}